\documentclass[12pt,onecolumn]{IEEEtran}

\usepackage{amsmath, amssymb, amsfonts}
\usepackage[amsmath,thmmarks,amsthm]{ntheorem}
\usepackage{enumerate}
\usepackage{setspace}
\usepackage{graphicx}
\usepackage{verbatim}
\usepackage{url}

\usepackage{algorithm}
\usepackage[noend]{algpseudocode}
\makeatletter
\def\BState{\State\hskip-\ALG@thistlm}
\makeatother

\def\urltilde{\kern -.15em\lower .7ex\hbox{\~{}}\kern .04em}
\def\urldot{\kern -.10em.\kern -.10em}
\def\urlhttp{http\kern -.10em\lower -.1ex\hbox{:}\kern -.12em\lower 0ex\hbox{/}\kern -.18em\lower 0ex\hbox{/}}

\newtheorem{theorem}{Theorem}
\newtheorem{proposition}[theorem]{Proposition}
\newtheorem{lemma}[theorem]{Lemma}
\newtheorem{corollary}[theorem]{Corollary}

\newtheorem{example}{Example}
\newtheorem{problem}{Problem}

\newcommand{\N}{\mathcal N}

\newcommand{\I}{\mathcal I}

\begin{document}
\title{Minimal Controllability of Conjunctive Boolean Networks is NP-Complete\thanks{Research supported in part by a research 
grant from the Israel Science Foundation~(ISF grant 410/15).}}
\author{Eyal Weiss,   Michael Margaliot and Guy Even\thanks{
The authors are with the School of Electrical Engineering-Systems, Tel Aviv University, Israel 69978.
 Corresponding author:
Prof. Michael Margaliot, School of Electrical Engineering-Systems and the Sagol School of Neuroscience, Tel Aviv University, Israel 69978.
Homepage: {\tt www.eng.tau.ac.il/\urltilde michaelm}
\;\;  Email: {\tt michaelm@post.tau.ac.il}  }  }
\maketitle
\begin{abstract}
Given a conjunctive Boolean network~(CBN) with~$n$ state-variables, we consider the problem of finding a minimal
 set of
state-variables to directly affect with an input so
that the resulting conjunctive Boolean control network~(CBCN) is controllable. 
We give a necessary and sufficient condition for controllability of a~CBCN;
an~$O(n^2)$-time algorithm for testing controllability; and prove  that nonetheless 
 the minimal  controllability problem for~CBNs is NP-hard. 
\end{abstract}
\section{Introduction}
Many modern networked systems include a large number of nodes (or state-variables).
Examples range from     the electric grid to  complex biological processes. If the system includes control inputs then
  a natural question  is whether the system is \emph{controllable}, that is, whether the
control authority is  powerful   
 enough to steer the system from any initial condition to any desired final condition.
Controllability is an important property of   control systems, and it
 plays a crucial role in many control problems, such as stabilization of
 unstable systems by feedback, and optimal control~\cite{sontag_book}.

 If the system is not controllable  (and in particular if there are no control inputs) then an important 
  problem    is  what is the \emph{minimal}
 number of    control inputs that should be added to the network so that it becomes controllable.
This calls for finding the key locations within the system such that controlling them
allows driving the entire system to any desired state.
This problem is interesting both theoretically and for  applications, as in many real-world 
systems it is indeed possible to 
add control actuators, but naturally this may be timely and costly, so minimizing the number of added controls is desirable.

Several recent papers studied     minimal controllability 
in   networks with a linear and time-invariant~(LTI) dynamics~(see, e.g.~\cite{Olsh_2014,slotine_nature} and the references therein).
In particular, 
Olshevsky~\cite{Olsh_2014} considered the following problem. Given the~$n$-dimensional 
 LTI system~$\dot x_i(t)=\sum_{j=1}^n a_{ij}x_j(t)$, $i=1,\dots,n$,
  determine  a \emph{minimal} set of indices~$\I \subseteq \{1,\dots,n\}$, 
such that the modified   system:
\begin{align*}
\dot x_i(t)&=\sum_{j=1}^n a_{ij}x_j(t)+u_i(t), &i\in\I,\\
\dot x_i(t)&=\sum_{j=1}^n a_{ij}x_j(t), & i \not \in\I,
\end{align*}
is controllable. Olshevsky~\cite{Olsh_2014} showed, using a reduction to the minimum hitting set problem,
that this problem 
 is NP-hard (in the number of state-variables~$n$).
 For a general survey on the computational complexity of various problems in systems and control theory, see~\cite{blondel}.

Boolean control networks~(BCNs) are discrete-time dynamical systems with Boolean state-variables and Boolean control inputs. A BCN without inputs is called a Boolean network~(BN).
BCNs date back to the early days of digital switching networks and neural network models with on-off type neurons. 
They have also been used to model many other important phenomena such as 
social networks (see, e.g.~\cite{soc_nets_bool,soc_nets_plos}), the spread of epidemics~\cite{bool_epidemcs}, etc.

More recently, BCNs have been extensively 
used to model biological processes and networks where the
possible set of states is assumed to  be finite (see, e.g.~\cite{born08,Helikar,faure06}).
For example, in gene regulation networks one may assume that each gene may be either ``on'' or ``off'' (i.e. expressed or not expressed).
Then the state of each gene can be modeled as a state-variable in a BN and the interactions between the genes 
(e.g., via the proteins that they encode) determine
the Boolean update function  for each state-variable. 

A BCN with  state-vector~$X(k)=\begin{bmatrix} X_1(k)&\dots&X_n(k)\end{bmatrix}'$
 is said to be controllable if for any pair of states~$a,b\in\{0,1\}^n$ there exists an integer~$N\geq 0$
and a  control sequence~$u(0),\dots, u(N-1)$ 
steering the state from~$X(0)=a$ to~$X(N)=b$.

A natural  representation of a BCN is via its graph of states~$G=(V,E)$, where the vertices~$V=\{1,\dots,2^n\}$
represent all the possible~$2^n$ states, and a directed edge~$e=(v_i \to v_j) \in E$
means that there is a control such that~$X(k)=v_i$ and~$X(k+1)=v_j$. Then clearly a BCN is 
controllable if and only if~$G$ is strongly connected~\cite{dima_cont}. 
Since testing strong connectivity of a digraph takes linear time in the number of its vertices and directed edges 
(see, e.g.~\cite{Tarjan72depthfirst}), one may expect  that verifying controllability of a general~BCN is intractable.

Akutsu et al.~\cite{connphard2007} showed, using a reduction to the 3SAT problem,
   that determining if there exists 
a control sequence steering   a~BCN between two given states
 is  NP-hard, and that this holds  even for~BCNs with 
 restricted network structures. This   implies in particular that verifying controllability is NP-hard.
Of course, it is still possible that the problems of verifying controllability and finding
 the minimal number of controls needed to make a BN controllable are tractable in some special
classes of Boolean networks.

An important special class of BNs  are those comprised of  
 nested canalyzing functions~(NCFs) \cite{kauffman_93}.
A Boolean function is called canalyzing
if there exists a certain value, called the canalyzing value,
such that any input with this value  uniquely determines the output of the function regardless of the other variables. 
For example, $0$ is a canalyzing value for the function~AND, as~$\text{AND}(0,x_1,\dots,x_k)=0$ for any~$x_i \in \{0,1\}$.
BNs with nested canalyzing functions are
  often used to
model   genetic networks~\cite{harris_2002,kauu_2003,kauu_2004}.

Here, we consider the subclass consisting of those NCFs that are
constructed only with the AND operator, the conjunctive functions.
A BN is  called a \emph{conjunctive Boolean network}~(CBN)  if   every
  update  function  includes only  AND operations, i.e., 
	the state-variables   satisfy an equation of the form:
\begin{equation}\label{eq:CBN}
X_i(k+1)=\prod_{j=1}^{n}(X_{j}(k))^{\epsilon_{ji}},\quad i=1,\dots,n, 
\end{equation}
where~$\epsilon_{ji}\in \{0,1\}$ for all~$i,j$. 

Recall that
a~BN is called a disjunctive Boolean network~(DBN) if   every
  update  function  includes only  OR operations.
	By applying De Morgan Law's, one can reduce~DBNs to~CBNs,
so all the results in this note  
hold also for~DBNs.

A CBN with~$n$ state-variables 
can be represented by its \emph{dependency graph} (or wiring diagram) that has~$n$ vertices corresponding to the  Boolean state-variables.
There is a directed edge~$(i \to j)$ if $X_i(k)$ appears in the update function of~$X_j(k+1)$. That is, the
dependency graph encodes the variable dependencies in the update functions.
We will assume from here  on  that none of the
update  functions is constant, so every
vertex  in the dependency graph  has  a positive in-degree.
In this case, there is a one-to-one correspondence between the~CBN and its dependency graph,
and this allows a graph-theoretic analysis of the~CBN.
 For example, the problem of  characterizing 
  all the periodic orbits of a CBN 
	with a strongly connected
dependency graph  has been solved in~\cite{Jarrah2010}, and the robustness of these orbits has been studied in~\cite{basar_CBN}.

In the context of modeling
  gene regulation, CBNs encode 
  synergistic regulation of a gene by several transcription factors~\cite{Jarrah2010}, and there
 is increasing
evidence that this type of mechanism is common in regulatory networks~\cite{Merika01051995,Gummow2006,Nguyen2006.0012}.

Here, we consider  the following problem. 
\begin{problem}\label{prob:mincon}
Given a CBN with~$n$ state-variables, suppose that for any~$i\in\{1,\dots, n\}$
we can replace the update function of~$X_i(k+1)$ 
by an independent  Boolean control~$U_i(k)$.   Determine 
 a  minimal\footnote{In computer science, this is usually called a 
 \emph{minimum} cardinality set rather than a minimal set. We follow  the terminology used in control theory.} 
set of indices~$\I \subseteq \{1,\dots,n\}$, 
such that the modified   system:
\begin{align}\label{eq:mform}
  X_i(k+1)&=U_i(k), &i\in\I,\nonumber
  \\
  X_i(k+1)&= \prod_{j=1}^{n}(X_{j}(k))^{\epsilon_{ji}},& i \not \in\I,
\end{align}
is controllable. 
\end{problem}
We refer to a BCN in the form~\eqref{eq:mform} as a \emph{conjunctive Boolean control network}~(CBCN).

Problem~\ref{prob:mincon} is important because it calls for finding a minimal set
of key variables in the~CBN such that controlling them makes the   system controllable. Of course, an efficient algorithm
for solving this problem must 
  encapsulate an efficient algorithm for testing  controllability   of a~CBCN. 
  
\begin{example}
Consider Problem~\ref{prob:mincon} for the~CBN
\begin{align*}
X_1(k+1)&=X_2(k),\\
X_2(k+1)&=X_1(k)X_2(k). 
\end{align*}
Suppose that we replace the update function for~$X_2(k)$ by a control~$U_2(k)$ so that the network becomes: 
\begin{align*}
X_1(k+1)&=X_2(k),\\
X_2(k+1)&=U_2(k). 
\end{align*}
This CBCN is clearly controllable. Indeed, 
given a desired final state~$s=\begin{bmatrix} s_1& s_2 \end{bmatrix}' \in\{0,1\}^2$, the control sequence~$U_2(0)=s_1$, $U_2(1)=s_2$, 
steers the CBCN from an arbitrary initial condition~$X_1(0),X_2(0)$
to~$\begin{bmatrix} X_1(2)&X_2(2)\end{bmatrix} ' =s$.  Thus, in this case a solution to Problem~\ref{prob:mincon} is 
to replace the update function of~$X_2$ by a control.
\end{example}

The main contributions of this note are: 
\begin{enumerate}
\item a necessary and sufficient condition for the controllability of a~CBCN;
\item a polynomial-time algorithm for
determining whether a~CBCN
 is controllable 
(more specifically the time complexity of this algorithm is~$O(n^2)$, where~$n$ is the number of state-variables in the BCN); 
\item a proof that Problem~\ref{prob:mincon} is NP-hard. 
\end{enumerate} 
Together, these results imply that
checking the controllability of a given~CBCN is ``easy'', yet 
  there   does not exist a polynomial-time algorithm for solving Problem~\ref{prob:mincon} (unless P=NP).

The next section 
reviews definitions and notations from graph theory 
that will be  used later on. Section~\ref{sec:main} describes our main results.   
 Section~\ref{sec:conclusion}  concludes and presents several directions for further research.
\section{Preliminaries}\label{sec:preliminaries}
Let $G=(V,E)$ be an undirected graph, where $V$ is the set of vertices,
 and $E$ is the set of edges. If 
 two vertices~$v_i$, $v_j$ are connected by an edge then we denote this edge 
  by~$e_{ij}$ or by~$(v_i,v_j)$, and  say that~$v_i$ and~$v_j$ are \emph{neighbors}. 
	The set of neighbors of~$v_i$ is denoted by $\N(v_i)$,  and the \emph{degree} of~$v_i$ is~$|\N(v_i)|$. 

A \emph{dominating set} for~$G$ is a subset $ D$ of $V$ such that every vertex in~$V\backslash D$ has at least one neighbor in~$ D$. 
 \begin{problem}[Dominating set problem]\label{prob:doms}
 Given a graph $G=(V,E)$ and a positive integer $k\leq|V|$, does there exist a dominating set $ D$ of $V$ with~$|D|\leq k$? 
\end{problem}

This is known to be an NP-complete decision problem (see, e.g.~\cite{Garey_and_Johnson}).

Let $G=(V,E)$ be a directed graph (digraph), with~$V$   the set of vertices, and $E$   the set of
 directed edges (arcs). Let~$e_{i\to j}$ (or   $(v_i\to v_j)$) denote
the arc from $v_i$ to $v_j$. When such an arc exists, we say that $v_i$ is an \emph{in-neighbor} of $v_j$, and $v_j$ as an \emph{out-neighbor} of~$v_i$. The set of in-neighbors and out-neighbors of a vertex $v_i$ is denoted by $\N_{in}(v_i)$ and $\N_{out}(v_i)$, respectively. The \emph{in-degree} and \emph{out-degree}
 of~$v_i$ are~$|\N_{in}(v_i)|$ and~$|\N_{out}(v_i)|$, respectively.

Let $v_i$ and $v_j$ be two vertices in $V$. A \emph{walk} from~$v_i$ to~$v_j$, denoted by $w_{ij}$, is a sequence:
 $v_{i_0}v_{i_1}\dots v_{i_q}$, with $v_{i_0}=v_i$, $v_{i_q}=v_j$,  
and~$e_{i_k\to i_{k+1}} \in E $ for all $k\in \{0,1,\dots q-1\}$. 
A \emph{simple path} is     a walk with pairwise distinct vertices. 
We say that $v_i$ is \emph{reachable} from $v_j$ if there exists a simple path from~$v_j$   to~$v_i$. 
A \emph{closed walk} is a walk   that starts and finishes at the same vertex. 
A closed walk is called  a \emph{cycle} if all the vertices in the walk are distinct, except for the start-vertex and the end-vertex.
A \emph{strongly connected digraph} is a digraph for which every vertex in the graph is reachable from any other vertex in the graph.

Recall that given the CBN  \eqref{eq:CBN}, the associated \emph{dependency graph} is a digraph $G=(V,E)$ with~$n$ vertices, such that $e_{i\to j} \in E$ if and only if $\epsilon_{ij}=1$.
A CBN is uniquely determined by its dependency graph, and for  this reason we interchangeably refer to the $i$th state-variable in the CBN and 
the $i$th vertex in its dependency graph.
We   extend the definition of a dependency graph to a CBCN in a natural way:
 upon replacing the update equation for~$X_i(k+1)$ to~$X_i(k+1)=U_i(k)$
 we   remove all the arcs pointing to~$v_i$,   introduce a new vertex $v_{U_i}$ for the new control input, 
and   add an arc~$e_{U_i\to i}$.

	An \emph{m-layer graph} is a digraph $G=(V,E)$ for which each layer~$L_k$, $k=1,\dots,m$, is a subset of $V$, 
	every vertex in $V$  belongs to a single    layer, and any arc $e_{i\to j} \in E$
	 is such that~$v_i \in L_k$ and $v_j \in L_{k+1}$, for some~$k \in \{1,2,\dots ,m-1\}$.

\section{Main Results}\label{sec:main}
\subsection{Complexity Analysis}
We begin by analyzing the computational complexity of Problem~\ref{prob:mincon}.
We will prove a hardness result for a~CBCN whose dependency graph is a 3-layer graph.
Our first result uses the  special  structure of this~CBCN   to provide a simple necessary and sufficient
 condition for controllability. We will later use this condition to relate controllability analysis for this~CBCN to the dominating
 set problem. 
\begin{lemma}\label{lemma:three_layer_cont}
Consider  a CBCN  with
a dependency graph   that is a 3-layer graph satisfying: every vertex in layer-1 is a control input to a vertex in layer-2,
and every vertex in layer-2 has {\bf its own}  control input (in layer-1). This CBCN
 is controllable if and only if every vertex in layer-3 has an in-neighbor (in layer-2) with out-degree equal to one. 
\end{lemma}

{\sl Proof of Lemma~\ref{lemma:three_layer_cont}.}
Assume that the CBCN is controllable. Seeking a contradiction, suppose that there exists a vertex~$v_i$
in layer~3 that 
 has no in-neighbor with out-degree one. Let~$X_i$ denote the corresponding state-variable. If~$v_i$ has zero in-neighbors 
then~$X_i(k)$ is constant, contradicting the assumption of controllability. 
Hence~$v_i$ must have at least one in-neighbor  and 
each of them with out-degree greater than one.
 From the controllability of the~CBCN, it follows that for any initial state~$X(0)$, there exists a time~$T\geq 0$
and a control sequence~$\{u(0),\dots,u(T-1)\}$ steering the CBCN
 to the final state:
\begin{align}\label{eq:fthis}
X_i(T)&=0, \nonumber 
\\
X_j(T)&=1 \text{ for all } j\neq i. 
\end{align}
  In other words, the state-variables corresponding to 
all the nodes in  layers 2 and 3, except for $X_i$, are one at time~$T$. But for~\eqref{eq:fthis}  to hold
at least one of the in-neighbors of~$v_i$ must be zero at time~$T-1$. Since  zero  is the canalyzing 
value  and every in-neighbor of~$v_i$ has out-degree greater than one,     there exists a
state variable $X_q$, $q\neq i$, such that~$X_q(T)=0$.   A contradiction. 
	
To prove the converse implication, suppose that every vertex in layer~3 of the CBCN
has an in-neighbor (in layer~2) with out-degree equal to one. We need to prove that the CBCN is controllable. 
Denote the nodes in layer~3 by~$w_1,\dots,w_q$.
For any~$i \in \{1,\dots,q\}$ the node~$w_i$ in layer~3 has an in-neighbor~$v_i$ in layer~2 
 such that~$v_i$ has out degree one.
Let~$p\geq q$ denote the number of nodes in layer~$2$, so that the nodes in layer~2 
are~$v_1,\dots,v_q,v_{q+1},\dots,v_p$, and their corresponding controls in
layer~1 are~$u_1,\dots,u_p$. 
 
Fix   arbitrary~$a\in\{0,1\}^p$ and~$b\in\{0,1\}^q$. We will show that for any initial condition
there exists a control sequence 
that steers the network  to the state~$v(4)=a$ and~$w(4)=b$, 
where~$v=[v_1 \dots v_p]', w=[w_1 \dots w_q]'$, 
thus proving controllability. 
In time steps~$0$ and~$1$, set  all the inputs to one.
Then~$v_i(2)=w_j(2)=1$ for all~$i,j$. 
Now let~$u_i(2)=b_i$ for all~$i\in\{1,\dots,q\}$, and~$u_i(2)=1$ for all~$i>q$. 
Then~$v_i(3)=b_i$ for all~$i\in\{1,\dots,q\}$, and~$v_i(3)=1$ for all~$i>q$. 
Finally, let~$u_i(3)=a_i$ for all~$i\in\{1,\dots,p\}$.
Then~$w_i(4)=v_i(3)=b_i$ for all~$i\in\{1,\dots,q\}$ (because $\text{AND}(1,z)=z$ for all~$z\in\{0,1\}$)
and~$v_i(4)=a_i$ for all~$i\in\{1,\dots,p\}$. This completes the proof of
 Lemma~\ref{lemma:three_layer_cont}.~\IEEEQED

We are now ready to present our main complexity result.
\begin{theorem}\label{theorm:reduction}
	The decision version of Problem~\ref{prob:mincon} is NP-hard.
\end{theorem}

{\sl Proof of Thm.~\ref{theorm:reduction}.}
Given an undirected graph $G=(V,E)$, and a positive integer~$k$,
consider Problem~\ref{prob:doms}.
We will show that we can solve this  instance of the dominating set problem
by solving a minimal controllability problem for
a 3-layer~CBCN with dependency  graph~$\tilde G=(\tilde V, \tilde E)$ constructed as follows. 
Define~$L_2:=E$, $L_3:=V$, and let~$\tilde{V}:= L_2\bigcup L_3$.
In other words,  $\tilde G$ 
 has~$|E|+|V|$ nodes. The nodes in~$L_3$ are denoted~$v_1,\dots,v_{s }$, where~$s:=|V|$ and
those in~$L_2$ are denoted by~$e_{uv}$.
The set~$\tilde E$ includes two sets of directed edges. First, every~$e_{uv} \in L_2$ 
induces an arc~$(e_{uv} \to e_{uv})$ (i.e., a self loop). 
Second, every  edge~$e \in E$  
induces two directed edges in~$\tilde E$: $(e_{uv}\to v)$ and~$(e_{uv} \to u)$. 
Thus,~$|\tilde E|= 3|E| $ (Example~\ref{exa:demo} below demonstrates this construction). 
Note that this construction is polynomial in~$|V|,|E|$.
	
Consider the CBN induced by~$\tilde G$ as a dependency graph, 
and the minimal controllability problem for this~CBN. 
A solution for this problem includes adding controls to some of the state-variables.
Since every node in~$L_2$ has a self-loop in~$\tilde G$, 
a control input must be added to each node~$e \in L_2$. 
Denoting by~$L_1$ the set of nodes corresponding to the control inputs 
added to the~CBN, it is evident that we obtained a 3-layer graph.
The solution to the controllability problem may also add control inputs to nodes in~$L_3$. 
Let~$Y\subseteq L_3$ denote the set of these nodes. 
Note that~$Y$ is also a set of nodes in the original graph~$G$. 
We require the following result. 

\begin{lemma}\label{lemma:dominating_set}
	The set~$Y$ is a minimum dominating set of~$G$.
\end{lemma}
{\sl Proof of Lemma~\ref{lemma:dominating_set}.}
Let~$\bar G$ denote the dependency graph of the controllable CBCN obtained by solving the minimal controllability problem described above. Define a new dependency graph~$\bar  G'$ by removing the nodes in~$Y$, their adjacent directed edges, and the nodes of the control inputs added to them in~$\bar G$. Then~$\bar G'$ is
 a 3-layer controllable CBCN: the first layer consists of the controls that
were added to the nodes in~$L_2$ (namely the nodes of~$L_1$), the second layer is composed of the nodes in~$L_2$, and the third is composed of the nodes in~$V\setminus Y$. 
It follows from  Lemma~\ref{lemma:three_layer_cont} that  for
 every vertex~$v \in (V\setminus  Y)$ there exists a vertex~$e  \in L_2$ such that~$e$ is an in-neighbor of~$v $ 
and~$e$ is \emph{not} an in-neighbor of any other vertex in~$V\setminus  Y$. Thus,~$e$ must be 
 an in-neighbor 
of a node in~$Y$. We conclude that~$Y$  is a 
dominating set of~$G$. Since 
the  construction solves the minimal controllability problem, it is clear that~$Y$  
is a minimum dominating set of~$G$.~\IEEEQED

We now consider two cases. If~$ |Y|\leq k$ then we found a dominating set with cardinality~$\leq k$ and thus 
the answer to Problem~\ref{prob:doms} for the given instance is yes. 
If~$|Y|>k$ then the argument above implies that there does not exist a dominating set with cardinality~$\leq k$ and thus
the answer to Problem~\ref{prob:doms} for the given instance is no. 
Since Problem~\ref{prob:doms}  is NP-hard,
this completes the proof of 
 Theorem~\ref{theorm:reduction}.~\IEEEQED. 

\begin{figure*}[t]
 \centering
 \scalebox{1}{\input{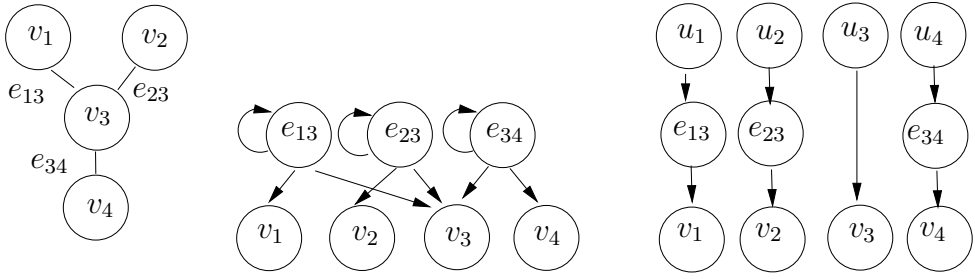}}
\caption{An example demonstrating the construction in the proof of Thm.~\ref{theorm:reduction}.
Left: a graph~$G$. The unique solution to the minimal dominating set problem for~$G$ is~$D=\{v_3\}$. 
Middle: the digraph~$\tilde G$ described in the proof of Thm.~\ref{theorm:reduction}. 
Right: the digraph~$\bar G$ 
describing the unique solution to the minimal controllability problem for~$\tilde G$ obtained by
adding  a control to each node~$e_{ij}$
and to the node~$v_3$ in~$\tilde G$.}\label{fig:const}
\end{figure*}

\begin{example}\label{exa:demo} 
Consider the graph~$G=(V,E)$ shown on the left of Fig.~\ref{fig:const}, with~$V=\{v_1,v_2,v_3,v_4\}$
and~$E=\{e_{13},e_{23},e_{34}\}$. The unique solution to the minimal dominating set for~$G$ is~$D=\{v_3\}$.
The construction in the proof above yields the directed graph~$\tilde G=(\tilde V,\tilde E)$ depicted in the middle of
Fig.~\ref{fig:const}. Note that this has two layers of nodes:~$L_2$ and~$L_3$ that include the edges and vertices in~$G$, respectively. 
The solution of the minimal controllability problem for~$\tilde G$ is depicted on the right of Fig.~\ref{fig:const}. Here~$Y=\{v_3\}$.
\end{example}


Our next goal is to present a polynomial-time algorithm
for determining if~\eqref{eq:mform} 
is controllable. To do this, we first derive graph-theoretic
necessary and sufficient conditions for the controllability of a~CBCN.
\subsection{Necessary and Sufficient Conditions for Controllability of a CBCN}
We begin by introducing several   definitions that will be used later on.
In a digraph, a \emph{source} is a node that has in-degree zero, and a \emph{sink} 
is a node that has out-degree zero. A digraph that does not contain cycles is called a \emph{directed acyclic graph}~(DAG).

We now introduce several 
 definitions referring  to the dependency graph of a~CBCN.
A node that represents a  state-variable~(SV)
	is called a \emph{simple node}.
 	A node that represents a control input is called a \emph{generator}.
 Note that a generator is always a source, and that
its set of out-neighbors always contains a single simple node (see~\eqref{eq:mform}).
 	A node that  represents an SV  that has an  added control input is called a \emph{directly controlled node}.
Note that a directly controlled node is also a simple node.
	A simple node with out-degree one, and without a self-loop, is called a \emph{channel}.
Note that the only out-neighbor of a channel is another simple node (since a generator is always a source).

We are now ready to derive necessary conditions for the controllability of a CBCN.
We will assume throughout that no~SV in the network has a constant updating function.
Indeed, in this case it is clear that the updating function of such an SV must be replaced by a control input. 
Under this assumption, 
 a simple node cannot be a source.
 
\begin{proposition}\label{prop:necessary_1}
	The dependency graph of a controllable~CBCN is acyclic.
\end{proposition}
 
{\sl Proof of Prop.~\ref{prop:necessary_1}.}
Consider a CBCN  with a cycle in its dependency graph. 
Every vertex in the   cycle corresponds to an SV (i.e., it is a simple node), as 
a generator is   a source, so it cannot be part of a cycle. 
Moreover, any simple node in the cycle
 is not a directly controlled node, since the only in-neighbor of a directly controlled node is a generator. 
This implies that if at time~$0$
  the SVs in the cycle are all zero then they can never be steered to the 
	a state where they are all one. Hence, the CBCN is not controllable.~\IEEEQED

It is natural to expect that in 
  a controllable~CBCN every~SV has the following property.
	There exists a path from a control input to the~SV that allows to set the~SV to zero (the canalyzing value),
	without 
affecting the other~SVs.
To make this precise, we say that a CBCN has \emph{Property~P} if
	every simple node in its dependency graph  
	contains in its set of in-neighbors either  
		a generator or  a channel.
The next result provides another necessary condition for controllability of a~CBCN.

\begin{proposition}\label{prop:necessary_2}
A controllable CBCN has  Property~P. 
\end{proposition}

{\sl Proof of Prop.~\ref{prop:necessary_2}.}
Seeking a contradiction, assume that
the~CBCN  is controllable and that there exists a   simple node~$v$ in its dependency graph  that does not contain a   generator  nor a channel  in its set of in-neighbors. This means that  there does not exist
a node~$w$  in the  graph such that~$v$ is the only simple node in the out-neighbors of~$w$.  Hence,
   the SV that corresponds to~$v$ cannot change its value to zero (the canalyzing value) without at 
	least one other~SV changing its value to zero  as well. 
	Consider two states:~$a$ corresponding to all SVs being zero,
	and~$b$ corresponding to all SVs being one except for~$v$ that is zero. 
	Since the CBCN is controllable it is possible to steer it from~$a$ to~$b$.
	This implies that $v$ has a self-loop, as it holds the value zero 
	 while the other SVs change their values. 
     Prop.~\ref{prop:necessary_1} implies that the~CBCN is not controllable.~\IEEEQED

Props.~\ref{prop:necessary_1} and~\ref{prop:necessary_2} provide two
 necessary conditions for the controllability of a~CBCN. The next result shows that
together they are also sufficient.
\begin{theorem}\label{theorm:cont_conditions}
	A CBCN is controllable if and only if its dependency graph
	  is a~DAG  and   satisfies Property~P.
		\end{theorem}

To prove this,  we introduce another definition and some auxiliary results. 
	A  \emph{controlled path} is an ordered non-empty set of nodes  in the dependency graph
	such that: the first element in the ordered set is a generator, and if 
		  the set contains more than one element, then for any~$i>1$ 
			the~$i$th node is a simple node, and is the \emph{only} element in the 
			set of out-neighbors of node~$i-1$.
	Controlled  paths with non-overlapping nodes are called  \emph{disjoint controlled paths}.

\begin{proposition}\label{prop:vertex_cover_by_IPs}
Consider a CBCN  with a dependency graph~$G$
	 that is a~DAG  and   satisfies Property~P.
		Then~$G$ can be decomposed 
	into disjoint controlled paths, such that every vertex in the graph  
	belongs to  a single controlled path (i.e., the union of the disjoint controlled paths forms a vertex cover of~$G$).
\end{proposition}
{\sl Proof of Prop.~\ref{prop:vertex_cover_by_IPs}}.
The proof is based on     Algorithm~1
detailed below  that  accepts such a graph~$G$  and terminates after each vertex in the graph 
  belongs to exactly one controlled path.  

\begin{algorithm} 
\caption{Decompose the  nodes in~$G$ into disjoint controlled paths}  
\begin{algorithmic}[1]
\While {there exists a simple node~$v\in G$ that is not included in any ordered set} \label{set:ini}
\State  $ \textit{cnode} \gets  v$ and $ \textit{cset} \gets  \{v\}$ 
 \If {$\N_{in}(\textit{cnode}) $ contains a channel} \label{set:ccha} 
	\State  pick a   channel~$u \in \N_{in}(\textit{cnode}) $  
	\If {$u$ does not belong to any ordered set} 
		\State insert $u$ to the ordered  set $ \textit{cset} $ just before~$\textit{cnode}$  
		\State    $ \textit{cnode} \gets  u$ 
		\State goto~\ref{set:ccha}   
	\Else
		\State let~$H$ denote the ordered set that contains~$u$\label{set:seth}
		\State merge    $\textit{cset} $ into~$H$ keeping the order between  any two adjacent elements
		\State goto \ref{set:ini}
	\EndIf 
\Else
	 \State pick a   generator~$u \in \N_{in}(\textit{cnode}) $   \label{line:used}
	 \If {$u$ does not belong to any ordered set} 
	         \State  insert it to $\textit{cset}$ just before~$\textit{cnode}$
						\State goto~\ref{set:ini}
						\Else
						\State goto~\ref{set:seth}
\EndIf \EndIf
\EndWhile 
\end{algorithmic}
\end{algorithm}

Note that the assumption that~$G$ has Property~P is used 
  in line~\ref{line:used} of the algorithm. 
The proof that Algorithm~1  
		terminates in a finite number of steps and
		divides   all the nodes in the graph   to a set of disjoint 
			controlled paths is straightforward and thus omitted. 
		The basic idea is that Property~P and the fact that the graph is a DAG 
		implies that we can ``go back'' 
		from any simple node through a chain of channels ending   with a generator,
		thus creating a  controlled path. 
  This completes the proof of Prop.~\ref{prop:vertex_cover_by_IPs}.~\IEEEQED

\begin{figure}[t]
 \centering
 \scalebox{1}{\input{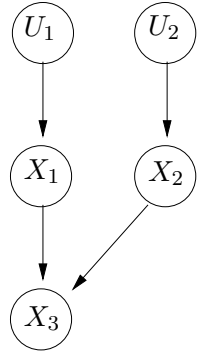}}
\caption{Dependency graph of the CBCN~\eqref{eq:amf}  .}\label{fig:sdg}
\end{figure}

\begin{example}
Consider the CBCN:
\begin{align}\label{eq:amf}
								X_1(k+1)&=U_1(k),\nonumber \\
								X_2(k+1)&=U_2(k),\nonumber \\
								X_3(k+1)&=X_1(k)X_2(k).
\end{align}
It is straightforward to verify 
that~\eqref{eq:amf} is controllable. 
Fig.~\ref{fig:sdg} depicts the dependency graph~$G$ of this CBCN.
Note that~$G$ is   a~DAG and that it satisfies Property~P. 
A decomposition into two disjoint controlled paths is~$C^1=\{U_1 \to X_1\to X_3\}$,~$C^2=\{ U_2\to X_2\}$. 
\end{example}

We can now prove Thm.~\ref{theorm:cont_conditions}.

{\sl Proof of Thm.~\ref{theorm:cont_conditions}.}
Consider a~CBCN satisfying the conditions stated in the Theorem. By 
 Prop.~\ref{prop:vertex_cover_by_IPs},   there exists a decomposition of its  dependency graph  
 into a set of~$m\geq 1$ disjoint controlled paths~$C^1,\dots,C^m$,  such that every vertex in the graph   belongs to a single 
  controlled path.   We assume from here on that~$m=2$ (the proof in  the general case is a straightforward generalization). 	
	
	Pick two states~$a,b\in\{0,1\}^n$. We prove that the~CBCN is controllable by providing a control sequence
	that steers the~CBCN from~$X(0)=a$ to~$X(T)=b$, in time~$T\geq 0$.  
	Since the paths provide a vertex cover, the desired state~$b$ can be decomposed into~$b^1$ and~$b^2$
	such that when the state is~$b$ the simple nodes in~$C^1$ [$C^2$] have state~$b^1$ [$b^2$]. 
	
	First, feed a control sequence with all ones to both generators
	until all the~SVs reach the value one at some time~$\tau \geq 0$. 
	Such a~$\tau$ exists, because by the properties of the paths there are no arcs between 
	simple nodes in~$C_1$ and~$C_2$, except perhaps an arc from the final node in one path, say~$C_1$,
	to the other path~$C_2$
	(and since the graph is a~DAG there is no arc from a node in~$C_2$ to a node in~$C_1$). 
	
	By adding  a chain of dummy control inputs at the beginning of the shorter path, if needed, 
	we may assume that~$|C^1|=|C^2|$. Now we may view each path as a shift register (with all SVs initiated to one) and it is
	straightforward to feed each path with a suitable sequence of 
	controls to obtain the desired states~$b^1$ and~$b^2$ at some time~$T\geq 0$. 
	Thus, the~CBCN is controllable.~\IEEEQED

Note that this proof also provides the sequence of controls needed to steer the~CBCN
 from~$a$ to~$b$, i.e.
it solves the \emph{control synthesis problem} (given    the decomposition into a set of disjoint controlled paths). Combining this with Prop.~\ref{prop:vertex_cover_by_IPs}
implies the following.
\begin{corollary}
A CBCN is controllable if and only if its dependency graph can be decomposed 
into a set of disjoint controlled paths.
\end{corollary}


Using the necessary and sufficient condition for controllability it is possible to 
derive an efficient algorithm for determining if a~CBCN is controllable. 
\subsection{An efficient    algorithm for determining controllability}
Algorithm~2 below   tests if a~CBCN is controllable. It is based on the condition in 
Thm.~\ref{theorm:cont_conditions}.

\begin{algorithm} 
\caption{Testing controllability of a~CBCN in the form~\eqref{eq:mform} with~$n$~SVs and~$q\leq n $ control inputs}  
\begin{algorithmic}[1]
\State generate the dependency graph~$G=(V,E)$ 
\If {$G$ is not a DAG} \label{set:dag}
\Return{(``not controllable'')}
\EndIf
\State create a list~$L$ of~$n$ bits \label{set:listl}
\State set all bits in~$L$ to 0
\For {all   nodes~$v \in V$}
						\If {$| \N_{out}( v ) | \not =1 $}
						\Return{(``not controllable'')}
						\Else
							\State $j \gets $ the element in $\N_{out}( v)$
							\State $L(j) \gets 1$
						\EndIf
\EndFor
{\bf endfor}
\If {all bits in~$L$ are 1} 
\State \Return{(``controllable'')}
\Else
\State \Return{(``not controllable'')}
\EndIf \label{set:listlend}
 \end{algorithmic}
\end{algorithm}

The input to   Algorithm~2 is a  CBCN with~$n$ SVs and~$q\leq n$ control inputs. 
The first step is to build the dependency graph~$G=(V,E)$. The complexity of this step  is~$O(n^2)$, as this
 requires
  going through~$n$ updating functions, and  each   function has at most~$n$ arguments. 
The resulting graph satisfies~$|V|=n+q\leq 2n $, and~$|E| \leq n^2$.

Checking if~$G$ is a DAG in line~\ref{set:dag}
can be done using a topological sort algorithm. The complexity
 is linear in~$|V|,|E|$ (see, e.g. \cite{Kahn62TopSort}), i.e. it is~$O(n^2)$.  

Lines~\ref{set:listl}-\ref{set:listlend} use the  list~$L$     to check Property~P, that is, to verify that
  the set of in-neighbors of every~$SV$ contains either a generator or a channel. 
This part has complexity~$O(n+q)=O(n)$.

The total time-complexity of the algorithm is thus~$O(n^2)$.  More precisely, the complexity of the algorithm is linear 
  in the
length of the description of the~CBCN, and the latter is~$O(n^2)$.

\section{Conclusions}\label{sec:conclusion}
Minimal controllability problems for dynamical systems
 are important both theoretically and for  real-world  applications where 
actuators can be added to control the~SVs. 
Here, we considered a minimal controllability problem for an important subclass of~BNs, namely,~CBNs.
Using a graph-theoretic approach we derived:  a 
 necessary and sufficient condition   for the~CBCN~\eqref{eq:mform} 
 to be controllable, and 
  a polynomial-time algorithm for testing controllability. We also  showed that 
the minimal controllability problem  
is~NP-hard. 

Our approach is based on the new concept of a controlled path and the decomposition of the dependency
 graph of the~CBCN 
into disjoint controlled paths. Roughly speaking, this corresponds to decomposing the~CBCN into a set of shift registers that are almost decoupled.

Recall that given a digraph~$G=(V,E)$ 
a \emph{path cover} is a set of directed paths such that every~$v\in V$ belongs to at least one path.
 A \emph{vertex-disjoint path cover} is a set of paths such that every~$v\in V$ belongs to exactly one path. 
The \emph{minimum path cover problem}~(MPCP) 
consists of finding a vertex-disjoint path cover 
having the minimal  number of paths. This problem has applications in software validation~\cite{Path_Cover1979}.
The~MPCP may seem closely related to the problem studied here, but this is not necessarily so. 
First, the~MPCP for a~DAG can be solved in polynomial time (see, e.g. \cite{graph_alg}).
Second, the solution of the~MPCP does not provide enough information on the 
controllability of a~CBCN. For this, we need the
more specific properties of controlled paths.

We believe that the notions introduced here will find more applications in other control-theoretic problems for~CBNs.
 An interesting direction for further research is to
 derive an  efficient   algorithm  
that is guaranteed to approximately solve the minimal controllability problem for~CBCNs, with 
a guaranteed approximation error in the number of needed control inputs.

\bibliographystyle{IEEEtran}
 \bibliography{eyal_bib}

\begin{thebibliography}{10}
\providecommand{\url}[1]{#1}
\csname url@rmstyle\endcsname
\providecommand{\newblock}{\relax}
\providecommand{\bibinfo}[2]{#2}
\providecommand\BIBentrySTDinterwordspacing{\spaceskip=0pt\relax}
\providecommand\BIBentryALTinterwordstretchfactor{4}
\providecommand\BIBentryALTinterwordspacing{\spaceskip=\fontdimen2\font plus
\BIBentryALTinterwordstretchfactor\fontdimen3\font minus
  \fontdimen4\font\relax}
\providecommand\BIBforeignlanguage[2]{{%
\expandafter\ifx\csname l@#1\endcsname\relax
\typeout{** WARNING: IEEEtran.bst: No hyphenation pattern has been}%
\typeout{** loaded for the language `#1'. Using the pattern for}%
\typeout{** the default language instead.}%
\else
\language=\csname l@#1\endcsname
\fi
#2}}

\bibitem{sontag_book}
E.~D. Sontag, \emph{Mathematical Control Theory: Deterministic Finite
  Dimensional Systems}, 2nd~ed.\hskip 1em plus 0.5em minus 0.4em\relax New
  York: Springer, 1998.

\bibitem{Olsh_2014}
A.~Olshevsky, ``Minimal controllability problems,'' \emph{IEEE Trans. Control
  of Network Systems}, vol.~1, no.~3, pp. 249--258, 2014.

\bibitem{slotine_nature}
Y.-Y. Liu, J.-J. Slotine, and A.-L. Barabasi, ``Controllability of complex
  networks,'' \emph{Nature}, vol. 473, pp. 167--173, 2011.

\bibitem{blondel}
V.~D. Blondel and J.~N. Tsitsiklis, ``{A survey of computational complexity
  results in systems and control},'' \emph{Automatica}, vol.~36, pp.
  1249--1274, 2000.

\bibitem{soc_nets_bool}
D.~G. Green, T.~G. Leishman, and S.~Sadedin, ``The emergence of social
  consensus in {Boolean} networks,'' in \emph{Proc. 2007 IEEE Symposium on
  Artificial Life (IEEE-ALife'07)}, Honolulu, Hawaii, 2007, pp. 402--408.

\bibitem{soc_nets_plos}
S.~Kochemazov and A.~Semenov, ``Using synchronous {Boolean} networks to model
  several phenomena of collective behavior,'' \emph{PLoS ONE}, vol.~9, no.~12,
  p. e115156, 2014.

\bibitem{bool_epidemcs}
A.~Kasyanov, L.~Kirkland, and M.~T. Mihaela, ``A spatial {SIRS} {Boolean}
  network model for the spread of {H5N1} avian influenza virus among poultry
  farms,'' in \emph{Proc. 5th Int. Workshop Computational Systems
  Biology~(WCSB)}, 2008, pp. 73--76.

\bibitem{born08}
S.~Bornholdt, ``Boolean network models of cellular regulation: prospects and
  limitations,'' \emph{J. R. Soc. Interface}, vol.~5, pp. S85--S94, 2008.

\bibitem{Helikar}
T.~Helikar, N.~Kochi, J.~Konvalina, and J.~Rogers, ``Boolean modeling of
  biochemical networks,'' \emph{The Open Bioinformatics Journal}, vol.~5, pp.
  16--25, 2011.

\bibitem{faure06}
A.~Faure, A.~Naldi, C.~Chaouiya, and D.~Thieffry, ``Dynamical analysis of a
  generic {B}oolean model for the control of the mammalian cell cycle,''
  \emph{Bioinformatics}, vol.~22, pp. e124--e131, 2006.

\bibitem{dima_cont}
D.~Laschov and M.~Margaliot, ``Controllability of {B}oolean control networks
  via the {P}erron-{F}robenius theory,'' \emph{Automatica}, vol.~48, pp.
  1218--1223, 2012.

\bibitem{Tarjan72depthfirst}
R.~Tarjan, ``Depth first search and linear graph algorithms,'' \emph{SIAM J.
  Computing}, vol.~1, no.~2, pp. 146--160, 1972.

\bibitem{connphard2007}
T.~Akutsu, M.~Hayashida, W.-K. Ching, and M.~K. Ng, ``Control of {Boolean}
  networks: Hardness results and algorithms for tree structured networks,''
  \emph{J. Theoretical Biology}, vol. 244, pp. 670--679, 2007.

\bibitem{kauffman_93}
S.~A. Kauffman, \emph{The Origins of order: Self Organization and Selection in
  Evolution}.\hskip 1em plus 0.5em minus 0.4em\relax Oxford University Press,
  1993.

\bibitem{harris_2002}
S.~E. Harris, B.~K. Sawhill, A.~Wuensche, and S.~Kauffman, ``A model of
  transcriptional regulatory networks based on biases in the observed
  regulation rules,'' \emph{Complexity}, vol.~7, no.~4, pp. 23--40, 2002.

\bibitem{kauu_2003}
S.~Kauffman, C.~Peterson, B.~Samuelsson, and C.~Troein, ``Random {Boolean}
  network models and the yeast transcriptional network,'' \emph{{Proceedings of
  the National Academy of Sciences}}, vol. 100, no.~25, pp. 14\,796--14\,799,
  2003.

\bibitem{kauu_2004}
S.~Kauffman, C.~Peterson, B.~Samuelssonand, and C.~Troein, ``Genetic networks
  with canalyzing {Boolean} rules are always stable,'' \emph{{Proceedings of
  the National Academy of Sciences}}, vol. 101, no.~49, pp. 17\,102--17\,107,
  2004.

\bibitem{Jarrah2010}
A.~S. Jarrah, R.~Laubenbacher, and A.~Veliz-Cuba, ``The dynamics of conjunctive
  and disjunctive {Boolean} network models,'' \emph{Bull. Math. Biology},
  vol.~72, no.~6, pp. 1425--1447, 2010.

\bibitem{basar_CBN}
\BIBentryALTinterwordspacing
Z.~Gao, X.~Chen, and T.~Basar, ``Stability structures of conjunctive {Boolean}
  networks,'' 2016. [Online]. Available:
  \url{https://arxiv.org/abs/1603.04415#}
\BIBentrySTDinterwordspacing

\bibitem{Merika01051995}
M.~Merika and S.~H. Orkin, ``Functional synergy and physical interactions of
  the erythroid transcription factor gata-1 with the {Kruppel} family proteins
  {Sp1} and {EKLF},'' \emph{Mol. Cell. Bio.}, vol.~15, no.~5, pp. 2437--2347,
  1995.

\bibitem{Gummow2006}
B.~M. Gummow, J.~O. Scheys, V.~R. Cancelli, and G.~D. Hammer, ``Reciprocal
  regulation of a glucocorticoid receptor-steroidogenic factor-1 transcription
  complex on the {Dax-1} promoter by glucocorticoids and adrenocorticotropic
  hormone in the adrenal cortex,'' \emph{Mol. Endocrinol.}, vol.~20, no.~11,
  pp. 2711--2723, 2006.

\bibitem{Nguyen2006.0012}
D.~H. Nguyen and P.~D{\textquoteright}haeseleer, ``Deciphering principles of
  transcription regulation in eukaryotic genomes,'' \emph{Molecular Systems
  Biology}, vol.~2, no.~1, 2006.

\bibitem{Garey_and_Johnson}
M.~R. Garey and D.~S. Johnson, \emph{Computers and Intractability: A Guide to
  the Theory of NP-Completeness}.\hskip 1em plus 0.5em minus 0.4em\relax San
  Francisco, CA: W. H. Freeman, 1979.

\bibitem{Kahn62TopSort}
A.~B. Kahn, ``Topological sorting of large networks,'' \emph{Communications of
  the ACM}, vol.~5, no.~11, pp. 558--562, 1962.

\bibitem{Path_Cover1979}
S.~C. Ntafos and S.~L. Hakimi, ``On path cover problems in digraphs and
  applications to program testing,'' \emph{IEEE Trans. Software Engineering},
  vol. SE-5, no.~5, pp. 520--529, 1979.

\bibitem{graph_alg}
S.~Even, \emph{Graph Algorithms}, 2nd~ed., G.~Even, Ed.\hskip 1em plus 0.5em
  minus 0.4em\relax New York, NY: Cambridge University Press, 2011.

\end{thebibliography}

\end{document}